\documentclass{aa}  
\usepackage{graphicx}
\usepackage{txfonts}

\begin{document}

   \title{Robust joint estimation of galaxy redshift and spectral templates using online dictionary learning}

   \author{S. Bryan\inst{1}\thanks{{\tt sean.a.bryan@asu.edu}}
          \and
          A.~Barekzai\inst{1}
          \and
          D.~Carter\inst{1}
          \and
          P.~Mauskopf\inst{1,2}
          \and
          J.~Mena\inst{1}
          \and
          D.~Rivera\inst{1}
          \and
          A.~Uriarte\inst{1}
          \and
          P.~Wang\inst{1}
          }

   \institute{School of Earth and Space Exploration, Arizona State University, Tempe, AZ 85287 USA
         \and
             Department of Physics, Arizona State University, Tempe, AZ 85281 USA
             }

   \date{Received October 30, 2024; accepted October 13, 2025}
 
  \abstract
   {We present a novel approach to analyzing astronomical spectral survey data using our non-linear extension of an online dictionary learning algorithm. Current and upcoming surveys such as SPHEREx will use spectral data to build a 3D map of the universe by estimating the redshifts of millions of galaxies. Existing algorithms rely on hand-curated external templates and have limited performance due to model mismatch error.}
   {We address these limitations by developing a new algorithm that jointly estimates both the underlying spectral features in common across the entire dataset, as well as the redshift of each galaxy.}
   {To do this, we significantly extend an existing online dictionary learning algorithm, and apply this approach to redshift estimation for the first time.}
   {Our new approach scales well to large datasets since we only process a single spectrum in memory at a time. Our algorithm performs better than a state-of-the-art existing algorithm when analyzing a mock SPHEREx dataset, achieving a normalized median absolute deviatio (NMAD) of 0.18\% and a catastrophic error rate of 0.40\% when analyzing noiseless data. Our algorithm also performs well over a wide range of signal to noise ratios (S/N), delivering sub-percent NMAD and catastrophic error above median S/N of 20. We released our algorithm publicly and it is available on github.}
   {}


   \maketitle

\section{Introduction}
\label{sec:introduction}
Estimating redshifts from spectra of galaxies is a fundamental capability that enables a broad range of studies in cosmology, astrophysics, and astronomy. Due to the expansion of the universe, the redshift of a target object is related to the distance between the observer and the target, which when combined with the observed ecliptic longitude and latitude of the target on the sky, enables current and future spectral surveys to map the 3D spatial distribution of galaxies across a large fraction of the observeable universe. Current and upcoming galaxy surveys such as SPHEREx \citep{crill20}, PAU \citep{benitez09} and others will measure millions of galaxies out to high redshift. These surveys have far more spectral bands with better spectral resolution. For example, SPHEREx has 102 spectral bands and PAU has 40, compared with 5 bands for SDSS \citep{fukugita96} in the previous generation of surveys. Existing approaches to estimate redshifts from galaxy spectra are highly performant \citep{cabanac02,ilbert06,brammer08}. However, these approaches rely on limited hand-curated external template data \citep{brown14} obtained from a small number of nearby galaxies. Galaxies evolve over time, and large samples of millions of galaxies will contain a broad diversity of processes that may not occur in the small number of well-studied nearby galaxies. For both of these reasons, large samples of high-resolution spectra of high-redshift galaxies will contain features not observed in nearby galaxies. This model mismatch error places a fundamental limit on the accuracy of redshifts obtained from existing data analysis approaches that rely on hand-curated spectral templates from nearby galaxies.

In this paper, we present results from a new algorithm that breaks through this fundamental limit by jointly estimating the spectral templates along with the redshifts of each galaxy in a survey. Our online algorithm analyzes one input spectrum at a time, and uses the difference between the data and the best fit to iteratively update the estimate of the spectral templates. By iterating on both a small subset of galaxies with known redshifts to initialize the algorithm, as well as iterating on the much larger set of galaxies without known redshifts, our algorithm converges towards a data-driven joint estimate of the redshifts and the spectral templates.

Our algorithm leverages and significantly extends an existing online dictionary learning \citep{mairal10} algorithm, and also builds on an earlier approach \citep{frontera-pons19} developed in the context of the SDSS/Rubin Observatory. Given a large dataset, dictionary learning robustly estimates a relatively small number of common modes, or atoms, that explain the fluctuations observed throughout that large dataset. Unlike a conventional principal component analysis \citep{jolliffe15} (PCA), and also unlike the {\tt photoz} template learning algorithm \citep{crenshaw20}, that accomplishes this task only if the entire dataset can be analyzed simultaneously, online dictionary learning algorithms are able to iteratively estimate and update these common atoms while analyzing only one item from the large dataset at a time. Our novel algorithm extends online dictionary learning by both estimating the atoms (i.e. the spectral templates), while also jointly estimating an unknown non-linear parameter (i.e. the redshift) of each item in our dataset.

\section{Estimating redshifts from spectra}

Redshifts derived from high spectral resolution ($R>1000$) spectral measurements currently provide the most accurate measurements of galaxy redshifts. Conventional instruments with this spectral resolution have significant limitations on the number of objects they can measure simultaneously, meaning that obtaining such spectroscopic redshifts for large numbers of faint galaxies is challenging. This motivates all-sky spectral surveys such as SPHEREx that instead measure galaxy redshifts using photometric data. This section reviews conventional approaches for estimating galaxy redshifts from photometry and evaluates their performance.

Current photometric redshift methods rely on comparing observed galaxy photometry to template spectra generated from models or observations of nearby galaxies. However, the complex physics involved in galaxy spectral energy distributions makes generating accurate template spectra difficult. The model mismatch between template spectra and real galaxy spectra is a major source of error in photometric redshift estimates. Section 2.1 reviews current template-fitting approaches for photometric redshift estimation. These methods approximate galaxy spectra using a limited dictionary of template spectra or even single template spectra. However, the resulting simplifications introduce model mismatch that degrades redshift accuracy. Section 2.2 defines the key performance metrics for redshift estimation, which are the normalized median absolute deviation (NMAD) and catastrophic error rate. These metrics characterize the accuracy and outlier rejection of photometric redshift fitting algorithms.

\subsection{Conventional redshift estimation approaches}

Current approaches \citep{brammer08,ilbert06} estimate the redshift of a single target galaxy by analyzing its spectrum $d_i$ measured at each of the $N_{bands}$ spectral bands of the instrument $\lambda_i^{obs}$. Determining the redshift from a single measured spectrum is a non-linear optimization problem of the form
\begin{equation}\label{conventional}
MSE(z,p) = \sum_{i=1}^{N_{bands}} (D(\lambda_i^{obs},z,p) - d_{i})^2,
\end{equation}
where $D(\lambda,p)$ is a function containing the underlying spectral features of any possible galaxy, $p$ is a set of parameters that control these features, and $MSE$ is the mean squared error between the model and the data. Varying the redshift $z$ and the parameters $p$ to minimize the $MSE$ would yield the maximum-likelihood estimate of $z$ from the input spectrum $d_i$. For clarity in presentation, here we assume that the error on each measurement $d_i$ is equal and follows a Gaussian probability distribution. Thus, minimizing the $MSE$ indeed is the maximum-likelihood estimator (MLE). The generalization to unequal Gaussian error bars is straightforward and performed by dividing the error terms by the variance of the error distribution.

The non-linear function $D$ is determined by all possible emission and absorption processes across a wide range of environments inside a galaxy, which are in general are non-linear with respect to their parameters $p$. A major challenge facing current approaches is that since $D$ is not known, current approaches approximate $D$ from observations of nearby galaxies and physics-based modeling. The resulting model mismatch error therefore places a significant limit on the performance of current approaches.

BAGPIPES \citep{carnall18}, CIGALE \citep{boquien19}, and FSPS \citep{conroy09} are codes that attempt to derive the non-linear functional form of $D$ from first principles and template data. These codes use Equation~\ref{conventional} to estimate the redshift and other parameters of measured galaxy spectra. They model emission from galaxies from the far-ultraviolet to the microwave regimes based on first principle models of the stellar populations, star formulation histories, dust attenuation, and other infrared/radio emission processes. Both BAGPIPES and CIGALE have a very large number of user parameters, leading to possible overfitting of the data. Also, first-principles models of galaxies do not yet capture the full observed complexity of these systems. Both of these effects, in addition to the extensive computing resources required for non-linear models, limit the applicability of this approach.

To make the analysis of large datasets feasible, the problem is often partially linearized. In EAZY \citep{brammer08}, the authors use data taken from a small set of nearby galaxies to construct a fixed spectral template dictionary $D_{l,m}$ sampled at a set of wavelengths $\lambda_m$ with $l$ counting among the 7 templates. EAZY finds the linear coefficient for each template, along with the overall redshift, to yield the best-fit redshift and galaxy spectrum. With the LePHARE \citep{ilbert06} algorithm, the problem is reduced further to consider only a single template $D(\lambda)$, and the algorithm only estimates an overall scaling of this template. LePHARE then repeats this process with different single templates, and co-adds the resulting likelihoods of all of the single templates at the end. LePHARE reports its estimated redshift along with the single template that most closely matched the data. This approach inherently comes with significant model mismatch error, since no galaxy follows any one template exactly.

An emerging new class of algorithms leverages neural networks \citep{pasquet19,dey22,stivaktakis20} to estimate the redshift from an input spectrum. These algorithms have good performance and use similar or fewer computational resources than existing algorithms that need to exhaustively search the entire redshift range. However, availability of on-sky training data is limited, necessitating the use of synthetic data to train the neural networks. First principles models or linear combinations of templates from nearby galaxies are used to generate this synthetic data, so neural network approaches also are susceptible to model mismatch error. Some progress is being made in this area \citep{dey22}, but current neural network algorithms are not able to present data to the user (such as a best-fit model of the spectrum, or a breakdown of which spectral features drive the estimate) to explain the redshift result.

As presented in Section~\ref{our_algorithm}, our novel algorithm addresses the model mismatch, overfitting, and explainability limitations of current approaches.

\subsection{Evaluating performance}

To evaluate the performance of photometric redshift fitting algorithms, one typically uses a set of photometric data of galaxies (either simulated or real) with known redshifts $z_{true}$ (already known for simulated data, or from spectroscopic follow-up for real data). This dataset is then fed to the algorithm under test which returns its estimated redshifts $z_{est}$ and calculates the spread of $\Delta z=z_{\text{est}}-z_{\text{true}}$, specifically, $\sigma (\Delta z)$
\begin{equation}
    \sigma\left(\Delta z\right)\equiv \text{std}\left(\frac{\Delta z}{1+z_{\text{true}}}\right).
\end{equation}
To avoid being affected by extreme outliers, we instead use Normalized Median Absolute Deviation ($\sigma_{\text{NMAD}}$) defined \citep{brammer08,hoaglin86} as
\begin{equation} \label{sigma_nmad}
    \sigma_{\text{NMAD}}\equiv 1.48\cdot \text{median}\left(\left|\frac{\Delta z - \text{median}(\Delta z)}{1+z_{\text{true}}}\right|\right).
\end{equation}
$\sigma_{\text{NMAD}}$ is directly comparable to the standard deviation, but a lot less sensitive to outliers since it uses the median and not mean to estimate the variance.

Another useful statistic to evaluate performance is the catastrophic error rate. This is the percentage of instances where the estimated value differs from the true value by a large amount. For redshift fitting, the fractional difference is typically \citep{brammer08} denoted $\eta$ and the catastrophic error rate is defined as the percentage of data with \citep{ilbert06}
\begin{equation}
    \eta\equiv \frac{\left|\Delta z\right|}{1+z}>0.15.
\end{equation}
The field is moving towards building sensor and analytics systems that can deliver $\sigma_{\text{NMAD}}$ of a fraction of a percent, and a catastrophic error rate of a few percent, for millions of galaxies.

\section{Joint redshift and template estimation algorithm}\label{our_algorithm}

We have developed a novel algorithm to simultaneously estimate the redshift and spectral templates from spectral survey data. We created an initial base dictionary that updates iteratively by comparing redshifts and dictionary templates to a new spectra. The dictionary is updated by calculating the redshift that minimizes error and adjusting the parameters for the templates to best fit the new spectra. This is repeated for each spectra, and the full dataset is also processed through the dictionary multiple times to improve the robustness of the estimation. This results in a sparse representation of the data and a dictionary of spectral templates that can be linearly combined to accurately represent the real data. The base dictionary is initialized with a similar process using a small subset of training data containing known redshifts and setting the number of expected spectral templates.

\subsection{Dictionary learning}

Stepping back, our goal is to estimate a relatively small number of underlying modes that explain the fluctuations observed in our large dataset. For modest datasets that do not have any additional nonlinear parameters such as redshift to estimate, Principal Component Analysis (PCA) \citep{jolliffe15,cabanac02} is often the preferred way to accomplish this. PCA is a linear transformation that maps data from a higher-dimension space to a lower-dimension space. PCA decomposes data into a combination of variables that represent linear functions of the original dataset while remaining uncorrelated and maximizing variance. There are many methods of achieving PCA, all with the goal of finding a linear combination to best represent the data. A standard way of decomposing is finding the eigenvalues and eigenvectors of the covariance matrix of the data \citep{jolliffe15}. For larger datasets that do not have any additional parameters like redshift to estimate, online dictionary learning algorithms yield similar results to a PCA but only require a single data element in memory at once. Dictionary learning algorithms do this by iteratively updating the estimate of the underlying modes as individual new data elements stream into the algorithm. 

Existing online dictionary learning algorithms \citep{mairal10,scikit-learn} work well for determining the dictionary atoms through processing data in subsets or mini-batches, but they do not allow for simultaneous estimation of redshift or other non-linear parameters. To overcome this limitation, we developed our novel algorithm to simultaneously estimate the underlying dictionary atoms and additional non-linear parameters.

\subsection{Our algorithm}

Our algorithm aims to use spectra $d_i$ measured at each of the $N_{bands}$ spectral bands of the instrument $\lambda_i^{obs}$ observed towards each of the $N_{gal}$ galaxies in our catalog to jointly estimate both the underlying spectral features $D(\lambda,p)$ and redshifts $z$ for each individual galaxy. Expressed in its full non-linear form, this is an optimization problem to minimize the scalar Mean Squared Error ($MSE$) across the entire catalog, which is
\begin{equation}\label{full_problem}
MSE^{cat.}(z_k,p,D) = \sum_{k=1}^{N_{gal}} \sum_{i=1}^{N_{bands}} (D(\lambda_i^{obs},z_k,p) - d_{i,k})^2.
\end{equation}
Note that unlike Equation~\ref{conventional}, here $MSE^{cat.}$ is also a function of the dictionary $D$, and sums over the entire catalog (not just a single spectrum as with Equation~\ref{conventional}). The general problem therefore is to vary the redshifts $z_k$, vary the best-fit parameters $p$, and also vary the functional form of the spectral features $D$, to find the global minimum $MSE$ across the entire dataset.

\begin{figure}
\centering
\includegraphics[width=0.5\textwidth]{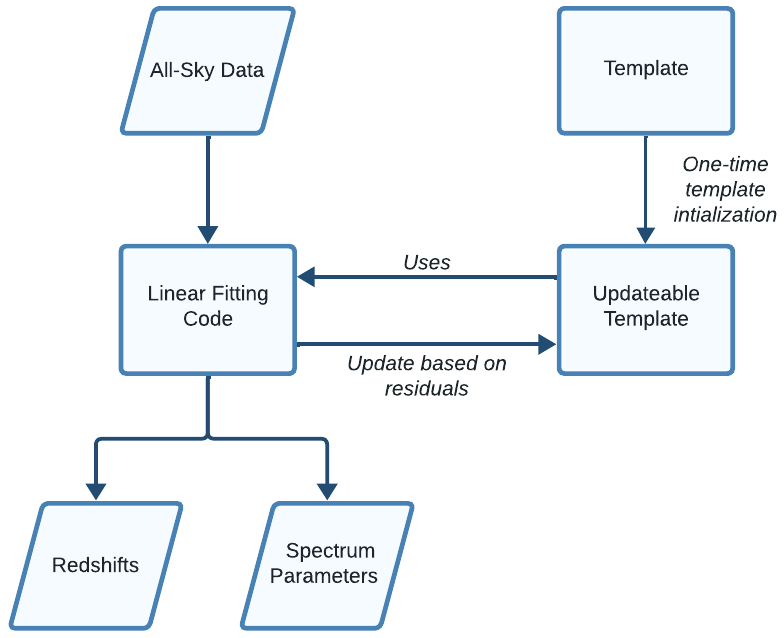}
\caption{High-level flowchart of our algorithm. We use an updateable template to continuously optimize the performance of the linear-fitting code based on the residuals between the model and the data.}
\label{flowchart}
\end{figure}

The full non-linear problem is not tractable, so we make the following two approximations. Since the functional form of $D$ is not known, our first approximation is to partially linearize the problem by modifying Equation~\ref{full_problem} to
\begin{equation}
MSE^{cat.}(z_k,p_{l,k},D_{l,m}) = \sum_{k=1}^{N_{gal}} \sum_{i=1}^{N_{bands}} \left(\sum_{l=1}^{N_{dict}} D^{k}_{l,i} p_{l,k} - d_{i,k}\right)^2,
\end{equation}
where $D^{k}_{l,i}$ is the subset of a main dictionary $D_{l,m}$ corresponding to the wavelength range indicated by the redshift $z_k$ and the spectral bands of the instrument $\lambda_i^{obs}$, and where the main dictionary $D_{l,m}$ has $N_{dict}$ independent rows. This reduces the optimization problem to finite dimensionality. In this form the problem is to minimize the $MSE$ by varying the redshifts $z_k$, and the now finite number of best-fit parameters $p_{l,k}$ and entries in the main dictionary $D_{l,m}$.

There is an ample number of datapoints in the large dataset to find an optimal solution. In the model problem we study in this paper, inspired by the SPHEREx wavelength coverage and expected sensitivity \citep{crill20}, we consider a mock dataset with $N_{gal} = 2,000$, $N_{bands} = 102$, and $N_{dict} = 7$ with a dictionary sampled at 600 points. This means that from our large dataset of $N_{gal} \times N_{bands} = 204,000$ data points, we jointly estimate $N_{dict} \times 600 = 4,200$ parameters in the dictionary, along with $N_{gal}$ redshifts and $N_{gal} \times N_{dict} = 14,000$ fit parameters. This is a total of $20,200$ fit parameters, less than a tenth of the $204,000$ data points in the catalog. This means that approximately 10 of the data points will ``average down'' to estimate each parameter, meaning a robust parameter estimate is possible. Note that while SPHEREx will characterize millions of galaxies, we consider $N_{gal} = 2000$ in this paper to demonstrate basic algorithm performance. Still, scaling up to a more realistic problem where $N_{gal} = 1$ million, the fit parameters would represent only $7.7\%$ of the data.

However, even this partially linearized form of the problem is not tractable. Using a fully non-linear optimization algorithm such as a Monte Carlo Markov Chain (MCMC) \citep{foreman-mackey13} or MultiNest \citep{feroz09} to explore a parameter space of $20,000+$ parameters is not feasible. This also does not scale well as the problem grows to larger catalogs of millions of galaxies.

To bring the problem into a computationally feasible form that does scale well, our second approximation is to use an online approach. Instead of analyzing the entire large dataset at once, our online algorithm only analyzes a single input spectrum $d_i$ at a time. As discussed in the next three subsections, our algorithm has three main phases: redshift estimation, dictionary update, and iteration. We repeat these steps iteratively across every spectrum in our large dataset until the dictionary converges. The algorithm then uses this converged (and now fixed) dictionary to estimate the redshift of every spectrum in our large dataset. This yields joint estimates for the dictionary, parameters, and redshifts.

\subsubsection{Redshift estimation}
For a given trial redshift $z_j$ in a range from 0 to $z_{max}$, we determine the rest frame wavelengths $\lambda_i^{rest}$ that correspond to each of the spectral bands of our instrument,
\begin{equation}
\lambda_i^{rest} = \lambda_i^{obs} / (1 + z_j).
\end{equation}
The main spectral template dictionary $D_{l,m}$ is defined for each for rest frame wavelength $\lambda_m^{dict}$ (sampled somewhat more finely in wavelength than the spectral bands of the instrument) and has an index $l$ corresponding to each individual template in the dictionary. We then interpolate the main dictionary onto the observed spectral bands
\begin{equation}\label{interp}
D^{trial}_{l,i} = \mathtt{interp}(\lambda_i^{rest},\lambda_m^{dict},D_{l,m}),
\end{equation}
where $\mathtt{interp}(x,x_p,f_p)$ is an interpolation function that resamples an array $f_p$ sampled at $x_p$ onto $x$. For simplicity we use linear interpolation, a further refinement of our algorithm would be to make full use of the measured passbands of the instrument in this interpolation. We then use generalized linear fitting \citep{garcia99} to fit the data $d_i$ to this dictionary. The best-fit parameters $p_l$ are therefore
\begin{equation}\label{linfit}
p = (\mathsf{D}^{trial} \mathsf{D}^{trial\dagger})^{-1} \mathsf{D}^{trial} d^\dagger,
\end{equation}
where $\mathsf{D}^{trial}$ and $d$ are respectively $D^{trial}_{l,i}$ and $d_i$ in matrix notation. The best-fit model $m_{i,j}$ in turn is
\begin{equation}\label{param2model}
m_{i,j} = \sum_{l=1}^{N_{dict}} D^{trial}_{l,i} p_l,
\end{equation}
and the mean squared error of this spectrum between the model and data is
\begin{equation}
MSE_j = \sum_{i=1}^{N_{bands}} (m_{i,j} - d_i)^2.
\end{equation}

We repeat this combination of interpolation and linear fitting for each trial redshift $z_j$ and calculate its corresponding $MSE_j$. We then note which of these trial redshifts has the lowest (i.e. best) $MSE$ and report this as our best fit redshift $z_{est}$ along with the corresponding best-fit parameters $p_l$.

\subsubsection{Dictionary update}

Armed with the best-fit redshift $z_{est}$, we then use the data $d_i$ to update our dictionary. This update step forms the core of our novel algorithm.

At this point in our algorithm, the dictionary has not converged \citep{nr}, and therefore there is a significant residual between the best-fit model and the data. Rewriting Equation~\ref{param2model} in matrix notation, our dictionary $\mathsf{D^{unconverged}}$ differs from the true dictionary $\mathsf{D}$ by an error term $\delta\mathsf{D}$, and the model differs from the data by $\delta\mathsf{m} \equiv \mathsf{m} - \mathsf{d}$. Thus,
\begin{equation}
(\mathsf{D} + \delta\mathsf{D})\mathsf{p} = \mathsf{d} + \delta\mathsf{m}.
\end{equation}
Note that with a fully converged dictionary, we would have $\mathsf{D}\mathsf{p}=\mathsf{d}$ (i.e., a fully converged dictionary would model the data without error). Canceling terms yields
\begin{equation}
\delta\mathsf{D} \mathsf{p} = \delta\mathsf{m}.
\end{equation}
Returning to array notation and isolating the dictionary error $\delta D_{l,i}$ yields
\begin{equation}\label{fullstep}
\delta D_{l,i} = (m_i - d_i) / p_l.
\end{equation}
Subtracting this estimate of the dictionary error from the unconverged dictionary would immediately bring the dictionary into full convergence at least for this individual data spectrum $d_i$.

In practice, applying repeated corrections to the dictionary in the form of Equation~\ref{fullstep} is unstable. Fully applying a dictionary correction derived only from a single item in the dataset also overfits the dictionary in some sense, causing subsequent iterations to oscillate instead of gradually accumulating updates across iterations. To improve stability, we first scale down the update by a learning rate $\epsilon$. We empirically find $\epsilon=0.01$ yields good convergence for our problem. To further improve the stability, we also experimented with other scalings of the parameters and residuals, and empirically we selected the following dictionary update rule,
\begin{equation}\label{empirical}
\delta D_{l,i}^{empirical} = \epsilon \times (m_i - d_i) \frac{p_l}{\sqrt{\sum_l p_l^2}},
\end{equation}
since we find it has good convergence properties for our problem. A similar update rule to Equation~\ref{empirical} appears in the literature \citep{mairal10}, but the literature equation lacks our learning rate term and also lacks the square root in the parameter normalization.

\subsubsection{Iteration}

\begin{figure}
\centering
\includegraphics[width=3.5in]{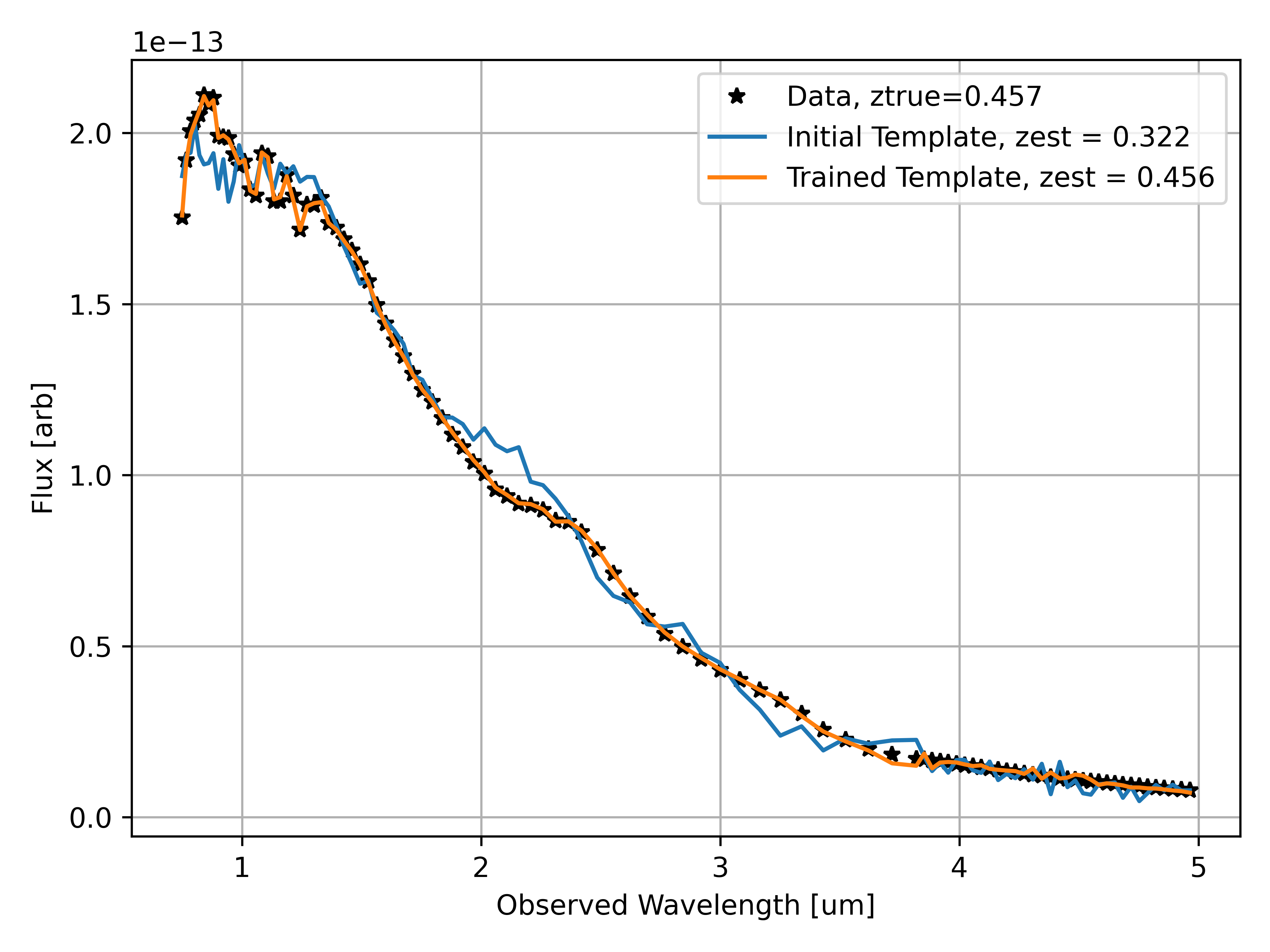}
\caption{Spectral fitting performance of a single representative spectrum using our algorithm. When the spectral dictionary is initialized, it has a poor fit (blue line) to the data (black points). When our algorithm iterates to convergence to estimate the dictionary, the best fit (orange line) is excellent and the best fit redshift ($z_{est} = 0.456$) very closely matches the ground truth ($z_{true} = 0.457$) even though the initial untrained dictionary did not estimate the correct redshift before training ($z_{init} = 0.322$).}
\label{spectrum_fitting}
\end{figure}

After estimating the redshift of the one spectrum in the large dataset, and applying the resulting dictionary update, we repeat this process for each other spectrum across the entire large dataset. We start by initializing the main dictionary $D_{l,i}$ with random values, that is for all $i$,
\begin{equation}
D_{l,i}^{init} = \mathcal{N}(\sigma_l,0),
\end{equation}
where $\mathcal{N}(\sigma,\mu)$ is a random number generated from a Gaussian distribution with mean $\mu$ and standard deviation $\sigma$. Since our dictionary update rule in Equation~\ref{empirical} scales dictionary updates according to the fit parameters, we initialize the dictionary with different amplitudes (i.e. standard deviations $\sigma_l$) for each row of the dictionary enumerated by $l$. To improve convergence, we initialize one row of the dictionary with Template \#1 from EAZY \citep{brammer08}. We do allow our algorithm to update this row in the dictionary. Also, to improve robustness we add an additional row to the dictionary where $D_{N_{dict}+1,i} = 1$ to enable the linear fitting algorithm in Equation~\ref{linfit} to estimate and account for the mean value of the spectrum. This row is not modified by the dictionary update algorithm.

To initialize the iteration process, we first iterate over the small subset of galaxies in our large dataset that have known redshifts (assumed to be 50 out of 2000 in our case). For these galaxies, we use the known redshift instead of estimating it from the data, and use this to guide the dictionary update step. After iterating over this subset, we then iterate over the much larger set of galaxies in the dataset with unknown redshifts. In turn, we repeat this entire process over the entire dataset several times to bring the dictionary to convergence.

Having the converged dictionary, we make a final pass through the entire large dataset (both the subset with known redshifts, and the set with unknown redshifts) to produce our final estimates of the redshifts using the fixed converged dictionary. The algorithm then outputs the dictionary and redshifts for presentation to the user.

\section{Performance}

\begin{figure}
\centering
\includegraphics[width=3.5in]{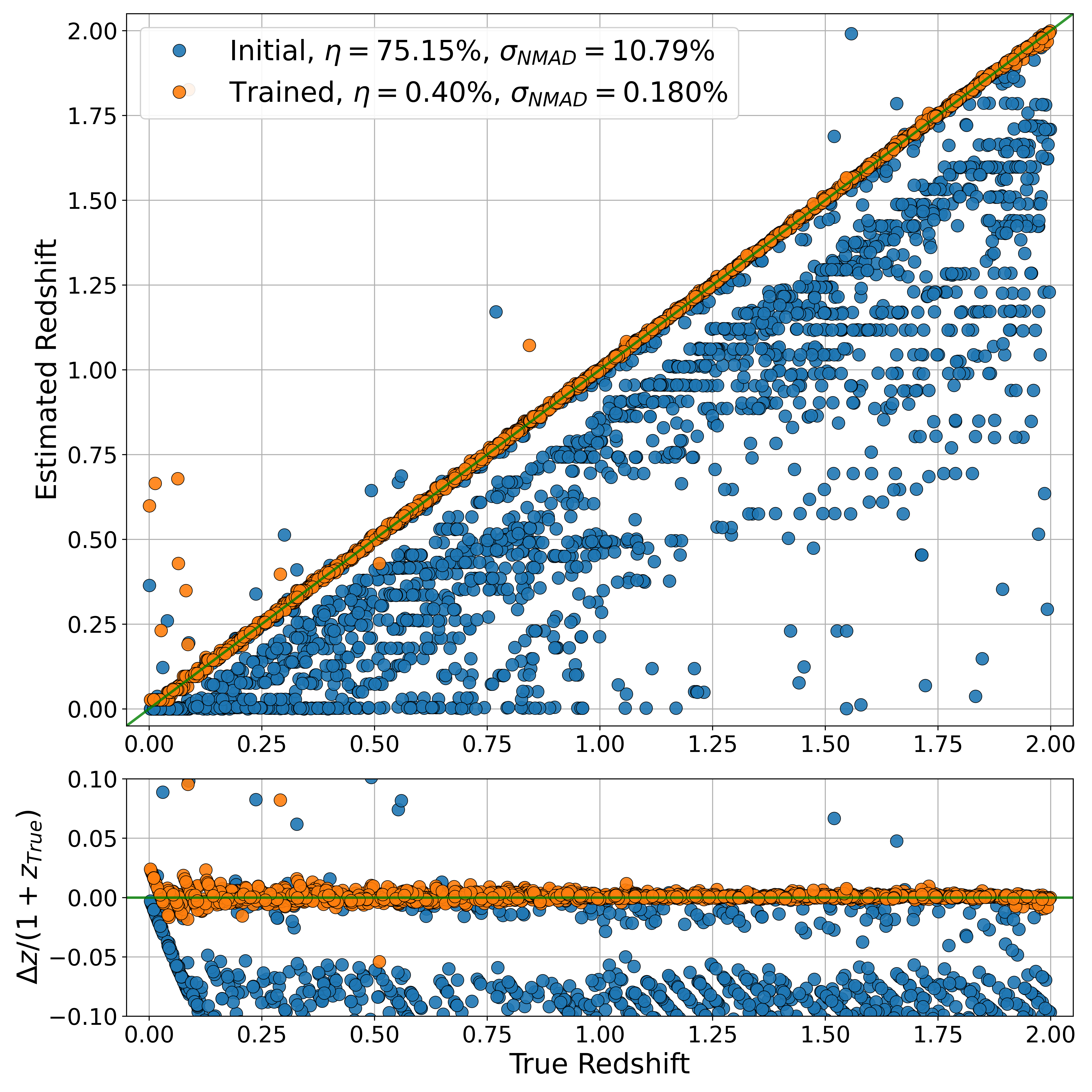}
\caption{Redshift reconstruction performance of our algorithm. Our algorithm has excellent catastrophic error (top panel, orange points) performance as well as scatter (bottom panel, orange points) performance after it learns the dictionary from the data. For comparison, when fitting the data with the initial dictionary before learning from the data, the performance is naturally poor (blue points).}
\label{performance}
\end{figure}

We created a simple mock dataset to evaluate the performance of our algorithm. For comparison, we also analyzed this dataset with the EAZY \citep{brammer08} algorithm. Our algorithm outperforms EAZY on this dataset, and is robust to noise. This result paves the way for our algorithm to analyze more complex datasets in the future, offering improved performance and mitigating model mismatch error.

\begin{figure*}
\centering
\includegraphics[width=7in]{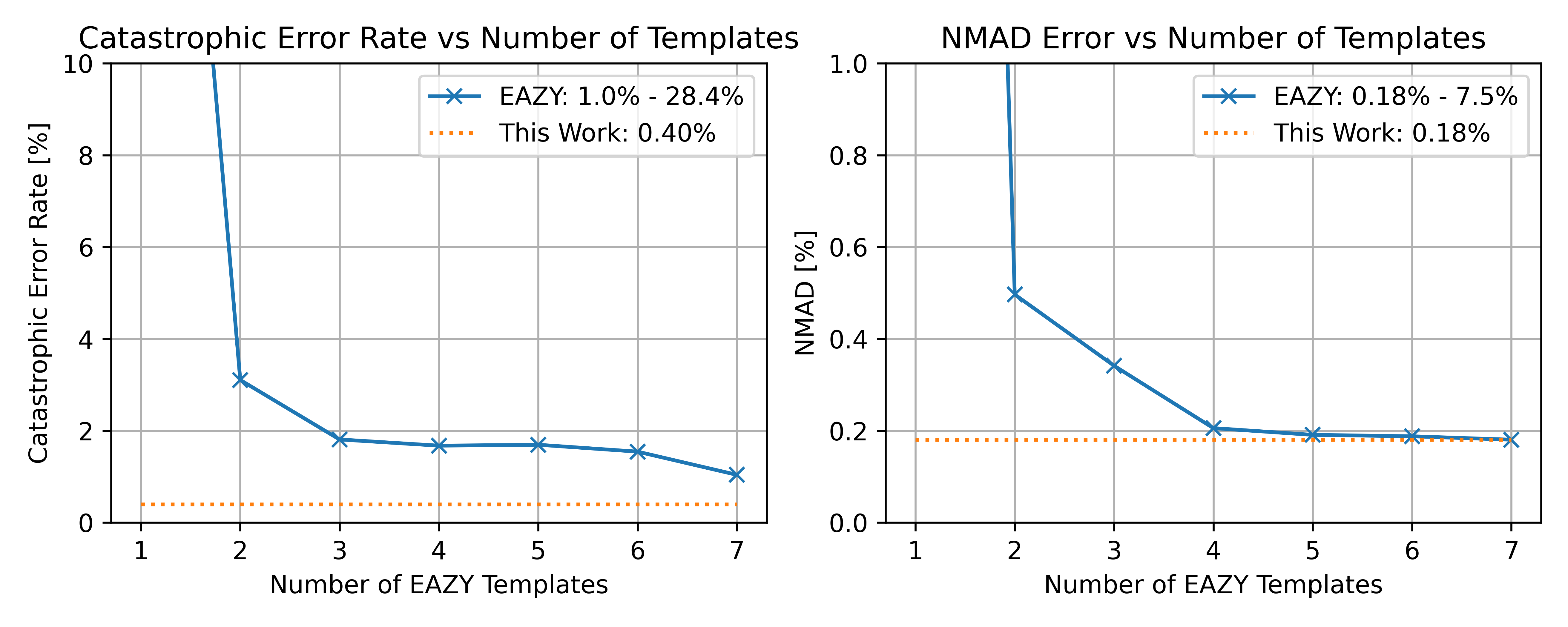}
\caption{Effect of template data denial on redshift reconstruction performance. When the EAZY \citep{brammer08} algorithm is given (a) fixed external template(s), the catastrophic error rate (left panel, blue line) and $\sigma_{\texttt{NMAD}}$ (right panel, blue line) are a strong function of the number of templates given to the algorithm. In contrast, we initialized our algorithm with only one template, and it self-consistently estimates all of the other templates to converge to excellent catastrophic error and scatter performance (orange dotted line). Since our algorithm does not require any external template data, initializing with no information yields similar performance (0.85\% catastrophic error, $\sigma_{\texttt{NMAD}}$=0.16\%).}
\label{err_vs_templates}
\end{figure*}

Our mock dataset is generated by forming linear combinations of the 7 spectral template modes EAZY uses to fit data and estimate redshifts. We used these modes to fit a common database \citep{brown14} of 129 nearby galaxy spectra. We noted the resulting means and standard deviations of the fit parameters, and generated Gaussian-distributed random numbers with matching means and standard deviations. We then used these mock parameters to scale the EAZY templates, producing a large mock dataset with similar statistical properties to the original 129 measured spectra. We generated random redshifts between 0 and 2 for each spectrum in our mock dataset, and convolved the redshifted spectra with nominal SPHEREx \citep{crill20} bandpasses to generate a noiseless mock SPHEREx galaxy spectral catalog. Since by construction there is no model mismatch error between this mock dataset and EAZY, this represents best-case performance for EAZY. We enable the template combination feature in EAZY to allow it to freely combine templates to yield the best fit. To demonstrate basic algorithm performance, our mock catalog contains 2,000 galaxies. SPHEREx will actually characterize millions of galaxies, and our online algorithm scales well to larger catalogs.

Our algorithm converges to an excellent best fit of the spectrum and redshift of spectra in our mock catalog. We initialized our spectral dictionary with Template \#1 from EAZY, and our algorithm iterated over the data to estimate the rest of the spectral dictionary. As shown in Figure~\ref{spectrum_fitting}, the converged spectral dictionary yields a far better fit to the spectrum than analyzing the data with Template \#1 alone. Figure~\ref{performance} shows that our algorithm achieves excellent reconstruction of the redshifts of the galaxies in the mock catalog. The catastrophic error is 0.4\%, and $\sigma_{\texttt{NMAD}}$ is 0.18\%.

Crucially, our algorithm achieves this redshift estimate result with little external input. This is in contrast with EAZY and other algorithms that require external template data, and deliver worse performance if the processes in the data are not reflected in the template. To simulate this effect, we denied EAZY access to its full set of templates, then reanalyzed our mock catalog generated with all seven EAZY templates. This quantifies the model mismatch error that existing approaches may face when analyzing future galaxy surveys with a diversity of unknown processes driving the spectral features.

We performed initial trials where we denied EAZY one template at a time, and observed the resulting decrease in redshift estimation performance. This let us rank order the templates in order of greatest to least impact on estimation performance. As shown in Figure~\ref{err_vs_templates}, we denied EAZY access to templates in order from least to greatest impact, observing progressively worse and worse performance. In contrast, even when initialized with only one template, our algorithm learns the other templates from the data and delivers consistent redshift estimation performance. In an apples-to-apples comparison, when EAZY was given one template its catastrophic error was 28.4\% and its $\sigma_{\texttt{NMAD}}$ was 7.5\%, far worse than the performance of our algorithm when it is only given the same single template (0.40\% and 0.18\% respectively). Worse, if EAZY is given no template data it has no capability to measure redshifts. In contrast, our algorithm still delivers excellent performance (0.85\% catastrophic error and 0.16\% $\sigma_{\texttt{NMAD}}$) even when it is initialized with no external template data at all. Even in the best case, when EAZY was given its full set of templates its catastrophic error was 1.0\% and its $\sigma_{\texttt{NMAD}}$ was 0.18\%, slightly worse than our algorithm.

\begin{figure*}
\centering
\includegraphics[width=7.25in]{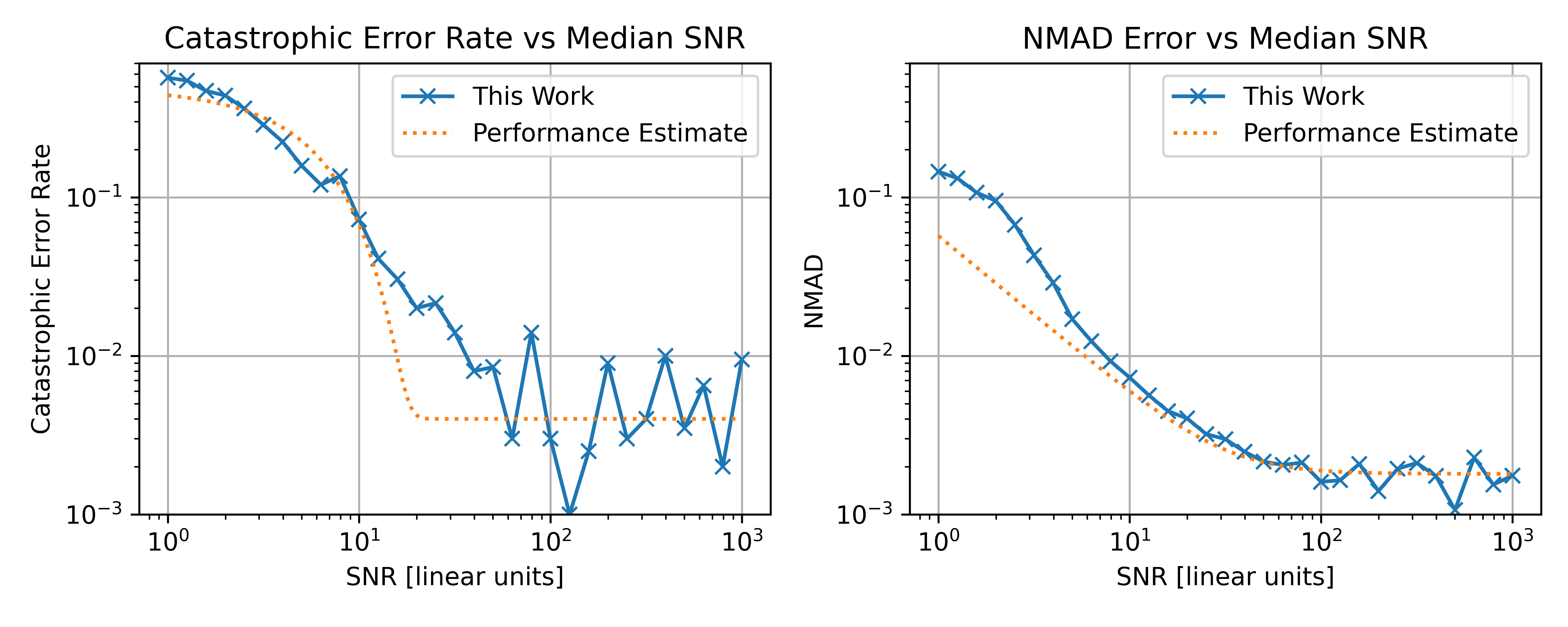}
\caption{Performance scaling of our algorithm with signal to noise ratio (S/N). Above an S/N of 20, our algorithm yields sub-percent catastrophic error rates $\eta$ (left panel) and $\sigma_{\texttt{NMAD}}$ (right panel). Our observed performance scales comparably to an approximate performance estimate discussed in the text.}
\label{err_vs_SNR}
\end{figure*}

We next conducted simulations to show that our algorithm is robust to noise. We added simulated Gaussian noise to the mock data, then reanalyzed the data with our algorithm and observed the resulting catastrophic error and $\sigma_{\texttt{NMAD}}$. As shown in Figure~\ref{err_vs_SNR}, above a median signal to noise ratio (S/N) of 20 our algorithm delivers sub-percent catastrophic error and $\sigma_{\texttt{NMAD}}$. Above S/N of 100 our algorithm delivers performance nearly indistinguishable from our noiseless results.

As a point of reference, we developed a rough estimate of how our performance nominally should scale with S/N. To construct this estimate for our problem, we consider a hypothetical spectral template that fully illuminates one of our passbands at $z=2$, but does not illuminate this passband at all at $z=0$. Thus, for unit S/N, this passband alone would estimate the redshift of the target with $\sigma_z^{nom} = \sqrt{\frac{2^2}{12}} \approx 0.577$ (using the formula for the standard deviation of a uniform deviate). Scaling up to the $N_{bands} = 102$ spectral bands of SPHEREx, the resulting performance estimate is
\begin{equation}
\sigma_z^{est}(\mathrm{S/N}) = \frac{\sigma_z^{nom}}{\mathrm{(S/N)} \sqrt{N_{bands}}}.
\end{equation}
Similarly for catastrophic error, we consider a hypothetical scenario in which the catastrophic error was only due to Gaussian fluctuations in redshift. Using the Gaussian error function (ERF) to calculate the rate of fluctuations outside the range $(-0.15,+0.15)$ yields a catastrophic error performance estimate of
\begin{equation}
\eta^{est}(\mathrm{S/N}) = \frac{1}{2} \left( 1 - ERF
\left(\frac{0.15 \mathrm{(S/N)}}{\sqrt{2}} \right) \right)
\end{equation}
As shown in Figure~\ref{err_vs_SNR}, our observed catastrophic error and $\sigma_{\texttt{NMAD}}$ performance scale similarly with this rough model when it is added to an observed error floor.

In the presence of data denial and noise, our algorithm has better performance than the state-of-the-art EAZY algorithm even when analyzing a mock dataset generated with EAZY's own templates. These promising performance results pave the way for our algorithm to be used on more complex datasets in the future.

\section{Possible future algorithm improvements}

Motivated by the good performance presented in this paper, we are considering several possible further improvements to our algorithm. The literature has a broad range of online dictionary update algorithms \citep{mairal10,scikit-learn} and we are studying them as potential replacements for our simple approach in Equation~\ref{empirical}. Also, our algorithm does not currently have any convergence criteria, so we are studying metrics we could track to stop our algorithm early if it already converged. In addition, as shown in Equation~\ref{interp}, we currently use linear interpolation to apply redshifts to our spectral dictionary. In the future, we plan to update this to an approach that allows for the use of measured instrument passbands. (Our software to generate our mock spectra already does account for instrument passbands.) Finally, we plan to refine the equations throughout our algorithm to optimally treat the case of passbands with unequal noise.

\section{Conclusion}

Our novel algorithm addresses model mismatch error in existing redshift estimation algorithms while offering better performance even when analyzing a simple mock dataset. Our online approach scales well to large datasets by only analyzing one spectrum at a time. The algorithm is robust to noise. We plan to apply this algorithm to more complex simulated and real datasets in the future to take advantage of its improved performance.

\section{Data and source code availability}

We released the source code of our algorithm publicly at {\tt github.com/HyperspectralDictionaryLearning/} \newline {\tt BryanEtAl2023 }. This includes the core algorithm that makes Figures~\ref{spectrum_fitting} and \ref{performance} directly, and that delivers the performance presented in for Figures~\ref{err_vs_templates} and \ref{err_vs_SNR}. We also included the code that generated the mock redshifted SPHEREx-observed spectra with similar statistical properties to the Brown \citep{brown14} galaxy sample using the EAZY \citep{brammer08} templates as a basis.

\begin{acknowledgements}
We thank Olivier Dor\'e, Daniel Masters, James Bock, Richard Feder-Staehle, BoMee Lee, Yongjung Kim, and Yun Ting Chen (i.e. the SPHEREx Redshift Pipeline Team) for helpful discussions during this research. We thank Henry Gebhardt for reviewing the manuscript.

This research was carried out under a contract with the National Aeronautics and Space Administration (80NM0018D0004).
\end{acknowledgements}

\bibliographystyle{aa}
\bibliography{report} 

@article{brammer08,
doi = {10.1086/591786},
url = {https://dx.doi.org/10.1086/591786},
year = {2008},
month = {oct},
publisher = {},
volume = {686},
number = {2},
pages = {1503},
author = {Gabriel B. Brammer and Pieter G. van Dokkum and Paolo Coppi},
title = {EAZY: A Fast, Public Photometric Redshift Code},
journal = {The Astrophysical Journal}}

@article{mairal10,
author = {Mairal, Julien and Bach, Francis and Ponce, Jean and Sapiro, Guillermo},
title = {Online Learning for Matrix Factorization and Sparse Coding},
year = {2010},
issue_date = {3/1/2010},
publisher = {JMLR.org},
volume = {11},
issn = {1532-4435},
journal = {J. Mach. Learn. Res.},
month = {mar},
pages = {19–60},
numpages = {42}
}

@book{nr,
author = {Press, William H. and Teukolsky, Saul A. and Vetterling, William T. and Flannery, Brian P.},
title = {Numerical Recipes 3rd Edition: The Art of Scientific Computing},
year = {2007},
isbn = {0521880688},
publisher = {Cambridge University Press},
address = {USA},
edition = {3}
}

@article{crenshaw20,
doi = {10.3847/1538-3881/abb0e2},
url = {https://dx.doi.org/10.3847/1538-3881/abb0e2},
year = {2020},
month = {sep},
publisher = {The American Astronomical Society},
volume = {160},
number = {4},
pages = {191},
author = {Crenshaw, John Franklin and Connolly, Andrew J.},
title = {Learning Spectral Templates for Photometric Redshift Estimation from Broadband Photometry},
journal = {The Astronomical Journal}
}

@article{conroy09,
doi = {10.1088/0004-637X/699/1/486},
url = {https://dx.doi.org/10.1088/0004-637X/699/1/486},
year = {2009},
month = {jun},
publisher = {The American Astronomical Society},
volume = {699},
number = {1},
pages = {486},
author = {Conroy, Charlie and Gunn, James E. and White, Martin},
title = {THE PROPAGATION OF UNCERTAINTIES IN STELLAR POPULATION SYNTHESIS MODELING. I. THE RELEVANCE OF UNCERTAIN ASPECTS OF STELLAR EVOLUTION AND THE INITIAL MASS FUNCTION TO THE DERIVED PHYSICAL PROPERTIES OF GALAXIES},
journal = {The Astrophysical Journal}
}

@book{garcia99,
author = {Garcia, Alejandro L.},
title = {Numerical Methods for Physics (2nd Edition)},
year = {1999},
isbn = {0139067442},
publisher = {Prentice-Hall, Inc.},
address = {USA},
edition = {2nd}
}

@ARTICLE{stivaktakis20,
  author={Stivaktakis, Radamanthys and Tsagkatakis, Grigorios and Moraes, Bruno and Abdalla, Filipe and Starck, Jean-Luc and Tsakalides, Panagiotis},
  journal={IEEE Transactions on Big Data}, 
  title={Convolutional Neural Networks for Spectroscopic Redshift Estimation on Euclid Data}, 
  year={2020},
  volume={6},
  number={3},
  pages={460-476},
  doi={10.1109/TBDATA.2019.2934475}}

@article{ilbert06,
	author = {{Ilbert, O.} and {Arnouts, S.} and {McCracken, H. J.} and {Bolzonella, M.} and {Bertin, E.} and {Le F\`evre, O.} and {Mellier, Y.} and {Zamorani, G.} and {Pell\`o, R.} and {Iovino, A.} and {Tresse, L.} and {Le Brun, V.} and {Bottini, D.} and {Garilli, B.} and {Maccagni, D.} and {Picat, J. P.} and {Scaramella, R.} and {Scodeggio, M.} and {Vettolani, G.} and {Zanichelli, A.} and {Adami, C.} and {Bardelli, S.} and {Cappi, A.} and {Charlot, S.} and {Ciliegi, P.} and {Contini, T.} and {Cucciati, O.} and {Foucaud, S.} and {Franzetti, P.} and {Gavignaud, I.} and {Guzzo, L.} and {Marano, B.} and {Marinoni, C.} and {Mazure, A.} and {Meneux, B.} and {Merighi, R.} and {Paltani, S.} and {Pollo, A.} and {Pozzetti, L.} and {Radovich, M.} and {Zucca, E.} and {Bondi, M.} and {Bongiorno, A.} and {Busarello, G.} and {De La Torre, S.} and {Gregorini, L.} and {Lamareille, F.} and {Mathez, G.} and {Merluzzi, P.} and {Ripepi, V.} and {Rizzo, D.} and {Vergani, D.}},
	title = {Accurate photometric redshifts for the CFHT legacy survey calibrated using the VIMOS VLT deep survey},
	DOI= "10.1051/0004-6361:20065138",
	url= "https://doi.org/10.1051/0004-6361:20065138",
	journal = {A\&A},
	year = 2006,
	volume = 457,
	number = 3,
	pages = "841-856",
}

@article{carnall18,
    author = {Carnall, A C and McLure, R J and Dunlop, J S and Davé, R},
    title = "{Inferring the star formation histories of massive quiescent galaxies with bagpipes: evidence for multiple quenching mechanisms}",
    journal = {Monthly Notices of the Royal Astronomical Society},
    volume = {480},
    number = {4},
    pages = {4379-4401},
    year = {2018},
    month = {08},
    issn = {0035-8711},
    doi = {10.1093/mnras/sty2169},
    url = {https://doi.org/10.1093/mnras/sty2169},
    eprint = {https://academic.oup.com/mnras/article-pdf/480/4/4379/25539546/sty2169.pdf},
}

@article{boquien19,
	author = {{Boquien, M.} and {Burgarella, D.} and {Roehlly, Y.} and {Buat, V.} and {Ciesla, L.} and {Corre, D.} and {Inoue, A. K.} and {Salas, H.}},
	title = {CIGALE: a python Code Investigating GALaxy Emission},
	DOI= "10.1051/0004-6361/201834156",
	url= "https://doi.org/10.1051/0004-6361/201834156",
	journal = {A\&A},
	year = 2019,
	volume = 622,
	pages = "A103",
}

@article{cabanac02,
	author = {{Cabanac, R. A.} and {de Lapparent, V.} and {Hickson, P.}},
	title = {Classification and redshift estimation by principal   component analysis},
	DOI= "10.1051/0004-6361:20020665",
	url= "https://doi.org/10.1051/0004-6361:20020665",
	journal = {A\&A},
	year = 2002,
	volume = 389,
	number = 3,
	pages = "1090-1116",
}

@article{jolliffe15,
	author = {{Jolliffe, I. T.} and {Cadima, J.}},
	title = {Principal component analysis: a review and recent developments},
	DOI= "10.1098/rsta.2015.0202",
        url= "https://doi.org/10.1098/rsta.2015.0202",
	journal = {Philosophical Transactions of the Royal Society A},
	year = 2015,
	volume = 374,
	number = 2065,
	pages = "20150202",
}

@article{foreman-mackey13,
doi = {10.1086/670067},
url = {https://doi.org/10.1086/670067},
year = {2013},
month = {feb},
publisher = {University of Chicago Press},
volume = {125},
number = {925},
pages = {306},
author = {Daniel Foreman-Mackey and David W. Hogg and Dustin Lang and Jonathan Goodman},
title = {emcee: The MCMC Hammer},
journal = {Publications of the Astronomical Society of the Pacific},
}

@article{feroz09,
    author = {Feroz, F. and Hobson, M. P. and Bridges, M.},
    title = "{MultiNest: an efficient and robust Bayesian inference tool for cosmology and particle physics}",
    journal = {Monthly Notices of the Royal Astronomical Society},
    volume = {398},
    number = {4},
    pages = {1601-1614},
    year = {2009},
    month = {09},
    issn = {0035-8711},
    doi = {10.1111/j.1365-2966.2009.14548.x},
    url = {https://doi.org/10.1111/j.1365-2966.2009.14548.x},
    eprint = {https://academic.oup.com/mnras/article-pdf/398/4/1601/3039078/mnras0398-1601.pdf},
}

@article{brown14,
doi = {10.1088/0067-0049/212/2/18},
url = {https://dx.doi.org/10.1088/0067-0049/212/2/18},
year = {2014},
month = {may},
publisher = {The American Astronomical Society},
volume = {212},
number = {2},
pages = {18},
author = {Michael J. I. Brown and John Moustakas and J.-D. T. Smith and Elisabete da Cunha and T. H. Jarrett and Masatoshi Imanishi and Lee Armus and Bernhard R. Brandl and J. E. G. Peek},
title = {AN ATLAS OF GALAXY SPECTRAL ENERGY DISTRIBUTIONS FROM THE ULTRAVIOLET TO THE MID-INFRARED},
journal = {The Astrophysical Journal Supplement Series},
}

@article{scikit-learn,
 title={Scikit-learn: Machine Learning in {P}ython},
 author={Pedregosa, F. and Varoquaux, G. and Gramfort, A. and Michel, V.
         and Thirion, B. and Grisel, O. and Blondel, M. and Prettenhofer, P.
         and Weiss, R. and Dubourg, V. and Vanderplas, J. and Passos, A. and
         Cournapeau, D. and Brucher, M. and Perrot, M. and Duchesnay, E.},
 journal={Journal of Machine Learning Research},
 volume={12},
 pages={2825--2830},
 year={2011}
}

@article{hoaglin86,
author = { David C.   Hoaglin  and  Boris   Iglewicz  and  John W.   Tukey },
title = {Performance of Some Resistant Rules for Outlier Labeling},
journal = {Journal of the American Statistical Association},
volume = {81},
number = {396},
pages = {991-999},
year  = {1986},
publisher = {Taylor & Francis},
doi = {10.1080/01621459.1986.10478363},
URL = {
        https://doi.org/10.1080/01621459.1986.10478363
},
eprint = {
        https://doi.org/10.1080/01621459.1986.10478363
}}

@ARTICLE{benitez09,
       author = {{Ben{\'\i}tez}, N. and {Gazta{\~n}aga}, E. and {Miquel}, R. and {Castander}, F. and {Moles}, M. and {Crocce}, M. and {Fern{\'a}ndez-Soto}, A. and {Fosalba}, P. and {Ballesteros}, F. and {Campa}, J. and {Cardiel-Sas}, L. and {Castilla}, J. and {Crist{\'o}bal-Hornillos}, D. and {Delfino}, M. and {Fern{\'a}ndez}, E. and {Fern{\'a}ndez-Sopuerta}, C. and {Garc{\'\i}a-Bellido}, J. and {Lobo}, J.~A. and {Mart{\'\i}nez}, V.~J. and {Ortiz}, A. and {Pacheco}, A. and {Paredes}, S. and {Pons-Border{\'\i}a}, M.~J. and {S{\'a}nchez}, E. and {S{\'a}nchez}, S.~F. and {Varela}, J. and {de Vicente}, J.~F.},
        title = "{Measuring Baryon Acoustic Oscillations Along the Line of Sight with Photometric Redshifts: The PAU Survey}",
      journal = {Astrophysical Journal},
         year = 2009,
        month = jan,
       volume = {691},
       number = {1},
        pages = {241-260},
          doi = {10.1088/0004-637X/691/1/241},
archivePrefix = {arXiv},
       eprint = {0807.0535},
 primaryClass = {astro-ph},
       adsurl = {https://ui.adsabs.harvard.edu/abs/2009ApJ...691..241B}
}

@inproceedings{crill20,
author = {Brendan P. Crill and Michael Werner and Rachel Akeson and Matthew Ashby and Lindsey Bleem and James J. Bock and Sean Bryan and Jill Burnham and Joyce Byunh and Tzu-Ching Chang and Yi-Kuan Chiang and Walter Cook and Asantha Cooray and Andrew Davis and OIivier Dor{\'e} and C. Darren Dowell and Gregory Dubois-Felsmann and Tim Eifler and Andreas Faisst and Salman Habib and Chen Heinrich and Katrin Heitmann and Grigory Heaton and Christopher Hirata and Viktor Hristov and Howard Hui and Woong-Seob Jeong and Jae Hwan Kang and Branislav Kecman and J. Davy Kirkpatrick and Phillip M. Korngut and Elisabeth Krause and Bomee Lee and Carey Lisse and Daniel Masters and Philip Mauskopf and Gary Melnick and Hiromasa Miyasaka and Hooshang Nayyeri and Hien Nguyen and Karin {\"O}berg and Steve Padin and Roberta Paladini and Milad Pourrahmani and Jeonghyun Pyo and Roger Smith and Yong-Seong Song and Teresa Symons and Harry Teplitz and Volker Tolls and Steve Unwin and Rogier Windhorst and Yujin Yang and Michael Zemcov},
title = {{SPHEREx: NASA's near-infrared spectrophotometric all-sky survey}},
volume = {11443},
booktitle = {Space Telescopes and Instrumentation 2020: Optical, Infrared, and Millimeter Wave},
editor = {Makenzie Lystrup and Marshall D. Perrin and Natalie Batalha and Nicholas Siegler and Edward C. Tong},
organization = {International Society for Optics and Photonics},
publisher = {SPIE},
pages = {114430I},
year = {2020},
doi = {10.1117/12.2567224},
URL = {https://doi.org/10.1117/12.2567224}
}

@ARTICLE{fukugita96,
       author = {{Fukugita}, M. and {Ichikawa}, T. and {Gunn}, J.~E. and {Doi}, M. and {Shimasaku}, K. and {Schneider}, D.~P.},
        title = "{The Sloan Digital Sky Survey Photometric System}",
      journal = {Astronomical Journal},
         year = 1996,
        month = apr,
       volume = {111},
        pages = {1748},
          doi = {10.1086/117915},
       adsurl = {https://ui.adsabs.harvard.edu/abs/1996AJ....111.1748F}
}

@ARTICLE{pasquet19,
       author = {{Pasquet}, Johanna and {Bertin}, E. and {Treyer}, M. and {Arnouts}, S. and {Fouchez}, D.},
        title = "{Photometric redshifts from SDSS images using a convolutional neural network}",
      journal = {Astronomy and Astrophysics},
         year = 2019,
        month = jan,
       volume = {621},
          eid = {A26},
        pages = {A26},
          doi = {10.1051/0004-6361/201833617},
archivePrefix = {arXiv},
       eprint = {1806.06607},
 primaryClass = {astro-ph.IM},
       adsurl = {https://ui.adsabs.harvard.edu/abs/2019A&A...621A..26P}
}

@ARTICLE{dey22,
       author = {{Dey}, Biprateep and {Andrews}, Brett H. and {Newman}, Jeffrey A. and {Mao}, Yao-Yuan and {Rau}, Markus Michael and {Zhou}, Rongpu},
        title = "{Photometric redshifts from SDSS images with an interpretable deep capsule network}",
      journal = {Monthly Notices of the Royal Astronomical Society},
         year = 2022,
        month = oct,
       volume = {515},
       number = {4},
        pages = {5285-5305},
          doi = {10.1093/mnras/stac2105},
archivePrefix = {arXiv},
       eprint = {2112.03939},
 primaryClass = {astro-ph.IM},
       adsurl = {https://ui.adsabs.harvard.edu/abs/2022MNRAS.515.5285D}
}

@article{frontera-pons19,
	author = {{Frontera-Pons, J.} and {Sureau, F.} and {Moraes, B.} and {Bobin, J.} and {Abdalla, F. B.}},
	title = {Representation learning for automated spectroscopic redshift estimation},
	DOI= "10.1051/0004-6361/201834295",
	url= "https://doi.org/10.1051/0004-6361/201834295",
	journal = {A\&A},
	year = 2019,
	volume = 625,
	pages = "A73",
}

\end{document}